\documentclass[prl, 10pt, floatfix, twocolumn,superscriptaddress]{revtex4-1}
\usepackage{amsmath} 
\usepackage{graphicx} 
\usepackage{verbatim} 
\usepackage{color} 
\usepackage{subfigure} 
\usepackage{hyperref} 
\usepackage{sidecap}
\usepackage{color}

\begin{document}

\title{Ground-State Dirac Monopole}
\author{E. Ruokokoski}
\affiliation{Department of Applied Physics/COMP, Aalto University, P.O.~Box 14100, FI-00076 AALTO, Finland}
\author{V. Pietil\"a}
\affiliation{Physics Department, Harvard University, Cambridge, Massachusetts 02138, USA}
\author{M. M\"ott\"onen$^{1,}$}
\affiliation{Low Temperature Laboratory, Aalto University, P.O.~Box 13500, FI-00076 AALTO, Finland}


\begin{abstract}
We show theoretically that a monopole defect, analogous to the Dirac magnetic monopole, may exist as
the ground state of a dilute spin-1 Bose--Einstein condensate. 
The ground-state monopole is not attached to a single semi-infinite Dirac string, but forms a point where 
the circulation of a single vortex line is reversed.
Furthermore, the three-dimensional dynamics of this monopole
defect are studied after the magnetic field pinning the monopole is removed and the 
emergence of antimonopoles is observed. Our scheme is experimentally realizable with the 
present-day state of the art.
\end{abstract}

\maketitle

\emph{Introduction.---}Topologically nontrivial configurations of quantum and classical fields play
a fundamental role in the physics of various phase transitions ranging from 
superfluids to the early universe~\cite{Kosterlitz:1972,Kibble:1976}. In
particular, they can give rise to exotic quantum states relevant to
electromagnetism~\cite{Dirac}, elementary particles~\cite{Skyrme:1961},
grand unified theories~\cite{Uni}, and cosmology~\cite{Cosmo}. Given the
difficulties of making detailed observations on cosmological scales or probing
conventional matter on ultrashort length scales, experimental evidence
for many of these topological excitations is scarce and indirect at most.

Dilute ultracold Bose gases with a spin degree of freedom can host a variety of
topologically interesting structures, such as
coreless vortices, knotted textures, skyrmions, and several types of
monopoles~\cite{coreless, skyrmions, knot, Savage, Ruoste, Skyrmion2, Stoof:2001, Ville2, Nakahara:1990}. The high
controllability of these quantum systems and their observability using optical
imaging techniques allow, in principle, detailed experimental investigations
of the various defects. In practice, however, topologically nontrivial
configurations occur as excited states and their topological nature dictates that 
they cannot be easily created from a topologically trivial ground state. 
In particular, configurations relevant to tabletop experiments of exotic 
phenomena in high-energy physics and cosmology can be especially challenging 
to create in an atomic gas since they require intricate manipulation of the atomic cloud in 
all three spatial dimensions \cite{Isoshima:2000a, Ogawa:2002, Suominen, Choi:2011, Ville}.

A major step forward was taken recently when a robust method for
creating an analogue of the Dirac monopole was proposed~\cite{Ville}. 
The Dirac monopole~\cite{Dirac} is the simplest model for a magnetic
point charge which unlike its electric counterpart, has not been convincingly
observed. In Ref.~\cite{Ville}, the monopole defect was created into
the spin texture of a dilute Bose--Einstein condensate (BEC) by adiabatically
modifying external magnetic fields. The resulting monopole state is not
the ground state of the system because of its tendency to degrade due to 
dynamical instabilities associated with the Dirac string. In this Letter, 
we demonstrate that an analogue of the Dirac monopole in an atomic gas can exist as a 
ground state configuration. We find that in an experimentally feasible time-independent magnetic 
field the ground state of the BEC corresponds to a strong-field seeking state (SFSS) with a 
monopole defect. A similar defect is also found for the weak-field seeking state (WFSS). 
In contrast to the previous studies, we find that the minimum-energy monopole is not attached 
to a single Dirac string with two circulation quanta but manifests itself as a point where the 
circulation of a single-quantum vortex is reversed. A similar configuration may be obtained 
from the ideal Dirac monopole by applying a gauge transformation~\cite{Auerbach:1994}.
Our results show that the Dirac monopole is a physically viable concept and can represent a 
robust and long-lived state.

We consider a condensate of $^{87}{\textrm{Rb}}$ atoms with ferromagnetic spin--spin
interactions~\cite{Rbpara}. The total hyperfine spin of the constituent atoms
is $F=1$, and the order parameter is a three-component spinor
field. In the presence of a strong external magnetic field, the condensate
spin tends to align with the local field. Hence, external fields can be used
to imprint topologically nontrivial spin textures to the condensate. For topological 
reasons, the ferromagnetic phase cannot sustain stable isolated point-like 
defects~\cite{UedaBook}. 
This does not, however, completely exclude the existence of point defects, as they 
may appear as endpoints of quantized vortices. These defects are characterised by a geometric 
charge $Q_{3D}$ corresponding to the area on the unit sphere covered by the condensate 
spin as one encloses the point defect in the spatial coordinate space~\cite{Ville}.

\begin{figure}[!h]
\begin{minipage}[t]{1\linewidth}
\centering
\includegraphics[width=1\textwidth]{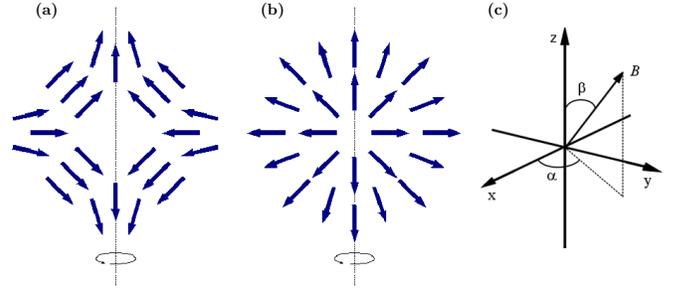}
\end{minipage}
\caption{\label{fig:textures}(a) Spin texture corresponding to the magnetic field in Eq.~(\ref{eq:field}).
(b) Vorticity $\boldsymbol{\Omega_{s}}$ corresponding to the spin texture in panel (a). Vector 
fields in panels (a) and (b) are symmetric
with respect to rotations about the axis depicted with the dashed line. (c) Illustration of the polar angles $\alpha$ and $\beta$ used to parametrize the local magnetic field $\boldsymbol{B}$. }
\end{figure}

Let us consider an external magnetic field which is a combination of two crossing quadrupole fields,
\begin{equation}\label{eq:field}
  \boldsymbol{B}\left(\boldsymbol{r}\right)=B^{\prime}_{1}\left(x\hat{e}_{x}+y\hat{e}_{y}
  \right)+B^{\prime}_{2}z\hat{e}_{z}.
\end{equation}
Since $\boldsymbol{B}$ is a monopole-free field, Maxwell's equation
$\nabla\cdot\boldsymbol{B}=0$ imposes a condition
$2B_{1}^{\prime}+B_{2}^{\prime}=0$.
The alignment of the hyperfine spin with the magnetic field in Eq.~(\ref{eq:field}) gives rise to a spin texture 
that is depicted in Fig.~\ref{fig:textures}(a) and has the geometric charge $Q_{3D}=1$.
The analogy to the Dirac monopole comes from the vorticity $\boldsymbol{\Omega}_{s}=\nabla
\times\boldsymbol{v}_{s}$, where $\boldsymbol{v}_{s}$ is the superfluid
velocity. The vorticity $\boldsymbol{\Omega}_{s}$ is equivalent to the magnetic 
field of a magnetic monopole and the Dirac monopole can be considered as a 
point source of the superfluid flow. Using the Mermin-Ho relation~\cite{Mermin-Ho}, 
the geometric charge $Q_{3D}$ of the spin texture can be related to the total 
vorticity in the system. Vorticity corresponding to the spin texture in Fig.~\ref{fig:textures}(a) 
takes almost everywhere the radially outward hedgehog form
\begin{equation}
\boldsymbol{\Omega}_{s} =\frac{\hbar}{mr'^{2}}\hat{\boldsymbol{e}}_{r'}
\end{equation}
in the scaled units $(x',y',z')=(x,y,2z)$~\cite{Ville}. This
establishes the analogy to the magnetic monopole proposed by Dirac~\cite{Dirac}.

\emph{Mean-field theory.---}In the zero-temperature limit, the stationary
states of the condensate are solutions to the time-independent
Gross--Pitaevskii (GP) equation~\cite{Ho}
\begin{equation}\label{eq:GP}
\mathcal{H}[\Psi]\Psi(\boldsymbol{r})=
\mu\Psi(\boldsymbol{r}),
\end{equation}
where the Hamiltonian for a spin-1 condensate is
\begin{equation}\label{eq:Hamilt}
\mathcal{H}[\Psi]=\hat{h}(\boldsymbol{r})+c_{0}|
\Psi\left(\boldsymbol{r}\right)|^{2}+c_{2}\Psi^{\dagger}
(\boldsymbol{r})\boldsymbol{\mathcal{F}}\Psi(\boldsymbol{r})
\cdot\boldsymbol{\mathcal{F}}.
\end{equation}
Here, $\boldsymbol{\mathcal{F}}=\left(\mathcal{F}_{x},\,\mathcal{F}_{y},\,
\mathcal{F}_{z}\right)^{T}$
is a vector of spin-1 matrices and $\hat{h}(\boldsymbol{r})$ is the
single-particle Hamiltonian given by
$\hat{h}(\boldsymbol{r})= -\hbar^{2}
\nabla^{2}/2m+V_{\textrm{opt}}(\boldsymbol{r})+g_{F}\mu_{B}\boldsymbol{B}
(\boldsymbol{r})\cdot\boldsymbol{\mathcal{F}}$,
where $g_{F}$ is the Land\'e $g$-factor and $\mu_{B}$
is the Bohr magneton. Unless otherwise mentioned, we consider a three-dimensional optical
trapping potential $V_{\textrm{opt}}(\boldsymbol{r})=m\omega_{r}^{2}r^{2}/2$.
The coupling constants are given by
$c_{0}=4\pi\hbar^{2}(a_{0}+2a_{2})/3m$ and $c_{2}=4\pi\hbar^{2}(a_{2}-a_{0})/3m$,
where $a_{f}$ is the $s$-wave scattering length corresponding to the channel
with total hyperfine spin $f$.
The Hamiltonian in Eq.~(\ref{eq:Hamilt}) corresponds to a free-energy functional of the form~\cite{Ho,Ohmi:1998}
\begin{equation} \label{eq:energy}
\begin{split}
E[\Psi]=\displaystyle\int_{}\textrm{d}\boldsymbol{r}\left[\frac{\hbar^{2}}{2m}\left|
\nabla\Psi(\boldsymbol{r})\right|^{2} + [V_{\textrm{opt}}(
\boldsymbol{r})-\mu]\left|\Psi(\boldsymbol{r})\right|^{2} \right.
\\
+ \left. \frac{c_{0}}{2}\left|\Psi(\boldsymbol{r})\right|^{4} +
\frac{c_{2}}{2}\left|\boldsymbol{S}(\boldsymbol{r})\right|^{2} +
g_{F}\mu_{B}\boldsymbol{B}
(\boldsymbol{r})\cdot\boldsymbol{S}(\boldsymbol{r})\right],
\end{split}
\end{equation}
where $\boldsymbol{S}=\Psi^{\dagger}\boldsymbol{\mathcal{F}}\Psi$.
The dynamics of the BEC are solved from the time-dependent GP equation,
$i\hbar\partial_{t}\Psi=\mathcal{H}[\Psi]\Psi$,
where $\mathcal{H}$ is given in Eq.~(\ref{eq:Hamilt}).

Let us write the stationary order parameter in the form $\Psi=\varphi\zeta$, 
where $\varphi$ is a scalar field and $\zeta$ is a unit-normalized spinor field. Assuming 
that the condensate spins align with the local field, the free energy of the system 
can be written as~\cite{Ho:1996}
\begin{equation}\label{eq:totH}
\begin{split}
E[\Psi]=\displaystyle\int_{}\textrm{d}\boldsymbol{r}\left\{\varphi^{\dagger}\left[\frac{1}{2m}\left(
-i\hbar\nabla+m\boldsymbol{v}_{s}\right)^{2} + g_{F}\mu_{B}B\left(\boldsymbol{r}\right)\right.\right.
\\
+\left.\left.\frac{\hbar^{2}}{2m}\left[\left|\nabla\zeta\right|^{2}+\left(\zeta^{\dagger}
\nabla\zeta\right)^{2}\right] +\mathcal{V}\right]\varphi\right\},
\end{split}
\end{equation}
where $\mathcal{V}$ is the interaction potential. Equation~(\ref{eq:totH}) is a Hamiltonian for a scalar particle 
with order parameter $\varphi$, and it is equivalent to the Hamiltonian for 
charged particles coupled to a vector potential $\boldsymbol{A}$. Hence, the equation of motion 
for the condensate is equivalent to that of charged particles in 
an electromagnetic field. The vector potential of the electromagnetic field corresponds to the 
superfluid velocity of the condensate if we set 
$m\boldsymbol{v}_{s}=q\boldsymbol{A}$~\cite{Ho:1996}.

For an optically confined BEC, minimization of the energy in Eq.~\eqref{eq:energy} results in a 
condensate in the SFSS which is lower in energy than the WFSS. We stress that only the SFSS monopole represents a robust ground-state
configuration. In order to study a realistic minimum-energy WFSS, we consider a purely magnetic trap provided by the two quadrupole fields in Eq.~(\ref{eq:field}).

We parametrize the magnetic field using polar
angles $\alpha$ and $\beta$ [Fig.~\ref{fig:textures}(c)] such that $\boldsymbol{B}\left(\boldsymbol{r}\right)=
|\boldsymbol{B}|\left(\sin\beta\cos\alpha,\,\sin\beta\sin\alpha,\,\cos\beta\right)^{T}$.
As the Land\'e $g$-factor is negative for spin-1 $^{87}{\textrm{Rb}}$ and we are considering the WFSS, the hyperfine spin
$\boldsymbol{S}$ is antiparallel with the local magnetic field.
In the eigenbasis of the spin operator $\hat{F}_{z}$, the order parameter can be expressed as
\begin{equation}{\label{eq:Zeeman}}
\left(\begin{array}{c}\Psi_{1}\\\Psi_{0}\\\Psi_{-1}\end{array}\right)={fe^{-i\gamma}}
\left(\begin{array}{c}e^{-i\alpha}\sin^{2}\left(\beta/2\right)\\
-\frac{1}{\sqrt{2}}\sin\beta\\e^{i\alpha}\cos^{2}\left(\beta/2\right)\end{array}\right)=
fe^{-i\gamma}\boldsymbol{\zeta},
\end{equation}
where $f$ is the amplitude of the order parameter and $\boldsymbol{\zeta}$ if fixed by 
Eq.~(\ref{eq:field}). 
We substitute the order parameter in Eq.~(\ref{eq:Zeeman}) into the stationary GP 
equation (\ref{eq:GP}) and denote $\Psi=f_{0}\boldsymbol{\zeta}$, 
where $fe^{-i\gamma}=f_{0}$. This gives a reduced GP equation which can be written in the dimensionless form as
\begin{equation}\label{eq:reduced}
\begin{split}
-\frac{1}{2}\left[\tilde{\nabla}^{2}+2\left(\boldsymbol{\zeta}^{\dagger}\tilde{\nabla}
\boldsymbol{\zeta}\right)\cdot\tilde{\nabla}+\left(\boldsymbol{\zeta}^{\dagger}
\tilde{\nabla}^{2}\boldsymbol{\zeta}\right)\right]\tilde{f}_{0}
\\
+\left|\tilde{\boldsymbol{B}}\right|\tilde{f_{0}}+\left(\tilde{c}_{0}+\tilde{c}_{2}\right)
\tilde{f_{0}}^{3}=\tilde{\mu}\tilde{f_{0}}.
\end{split}
\end{equation}
Here $\tilde{\mu}=\mu/\hbar\omega_{r}$, $\tilde{\boldsymbol{B}}=\mu_{B}\boldsymbol{B}/
\hbar\omega_{r}$, $\tilde{f_{0}}=a_{r}^{3/2}f_{0}$, $\tilde{c}_{i}=c_{i}N/(a_{r}^{3}\hbar\omega_{r})$
and  $a_{r}=\sqrt{\hbar/(m\omega_{r})}$. We normalize the complex-valued function $\tilde{f}_{0}$ 
to unity as $\int|\tilde{f}_0|^{2}\textrm{d}\tilde{r}=1$.
The minimum-energy density and phase distributions are solved from Eq.~(\ref{eq:reduced}) 
using the standard imaginary-time propagation
combined with finite-difference methods. The strong-field seeking ground state is found by minimizing
the free energy in Eq.~(\ref{eq:energy}) with full spin degrees of freedom using the successive overrelaxation method.
The temporal evolution of the monopole defect is solved from the time-dependent
GP equation using the split-operator and Crank--Nicolson methods.

\emph{Numerical results.---}The mass of a $^{87}\textrm{Rb}$ atom is taken to be
$m=1.44\times10^{-25}\textrm{ kg}$, the Land\'e $g$ factor is $g_{F}=-1/2$ and the total
number of atoms $N=8\times10^{4}$. For the coupling constants of $^{87}\textrm{Rb}$, we use
the value $c_{2}/c_{0}=-0.01$ and adopt $\tilde{c}_{0}=7500$. For $\omega_{r}=2\pi\times250\textrm{ Hz}$,
the dimensional values for the parameters are given by $B_{1}^{\prime}=0.1\,\textrm{T}/\textrm{m}$ 
for the simulations with the WFSS
and $B_{1}^{\prime}=-0.03\,\textrm{T}/\textrm{m}$ with the SFSS.
The volume considered in the simulation is $23\times23\times28$ in the units of $a_{r}^{3}$.
The computational grid consists of $141\times141\times161$ points.

For both SFSS and WFSS, the minimum-energy configuration corresponds to a monopole
defect associated with two vortices (Dirac strings) each carrying a single quantum of angular momentum. 
Both strings carry vorticity towards the monopole defect which lies in the zero point of the magnetic field. 
In Ref.~\cite{Ville}, the Dirac monopole is associated with only one vortex 
line that carries two quanta of vorticity. For energetic reasons, it is natural that this two-quantum vortex splits into two singly quantized vortices in the ground-state configuration~\cite{Huhtamaki:2006, Shin:2004a}.

\begin{figure}[!h]
\centering
\begin{minipage}[b]{1\linewidth}
\includegraphics[width=1\textwidth]{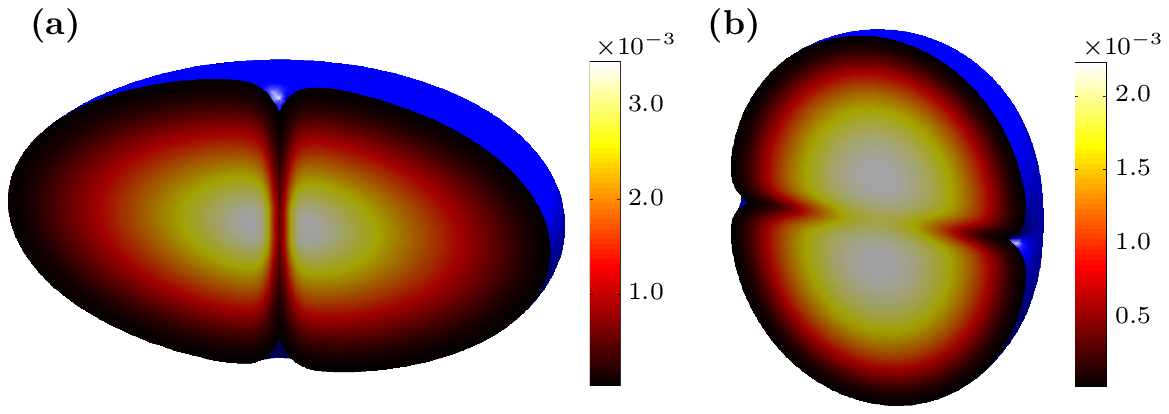}
\caption{\label{fig:dens}
Particle densities corresponding to (a) WFSS and (b) SFSS for $x>0$. 
The densities are given in the units of $N/a_{r}^3$.
The field of view is (a) $8\times16\times8$ and (b) $6\times12\times14$ in the units of $a_{r}^{3}$.\\
}
\end{minipage}
\begin{minipage}[b]{1\linewidth}
\includegraphics[width=1\textwidth]{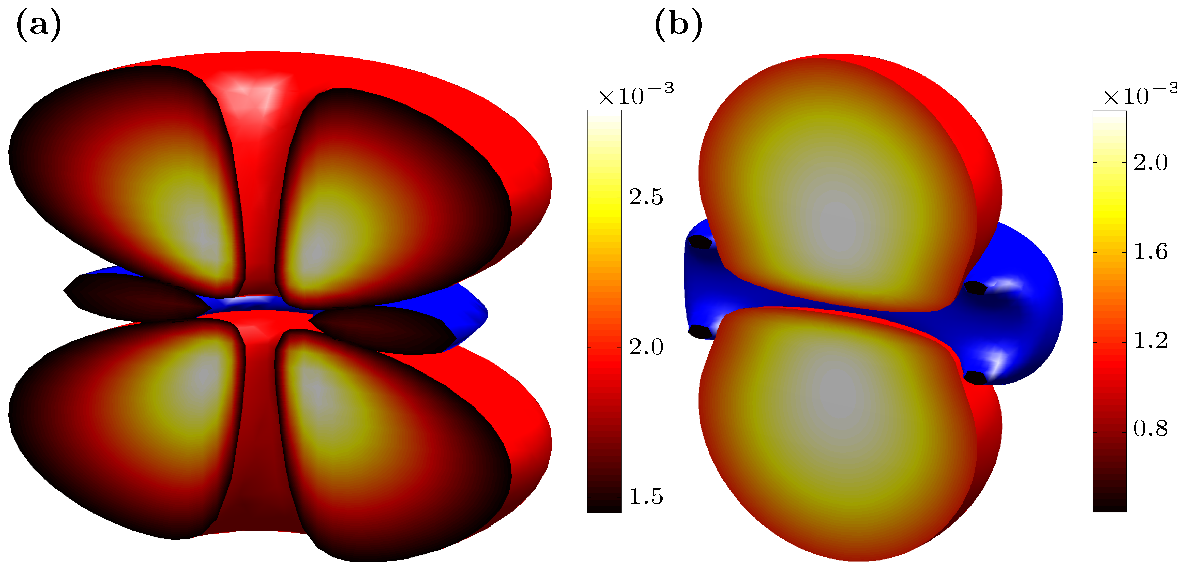}
\caption{\label{fig:spinor}
Densities of the three spinor components for (a) WFSS and (b) SFSS.
In both panels, the top segment corresponds to $\left|\Psi_{1}\right|^{2}$, the central segment to
$\left|\Psi_{0}\right|^{2}$, and the bottom segment to $\left|\Psi_{-1}\right|^{2}$. The colorbar scales are in the 
units of $N/a_{r}^3$. The segments are
bounded by density isosurfaces for spinor components for $x>0$. On the plane $x=0$, the isosurfaces
are capped with a density colormap for the corresponding spinor component.  In panel (a), all isosurfaces
correspond to density $1.4\times10^{-3}N/a_{r}^{3}$ and the singly quantized vortices are 
manifested as density depletion along the $z$-axis. In panel (b), the vortices lie on the 
$y$-axis and the isosurfaces correspond to densities
$1.2\times10^{-3}N/a_{r}^{3}$ for $\left|\Psi_{1}\right|^{2}$ and $\left|\Psi_{-1}\right|^{2}$ and
$4.5\times10^{-4}N/a_{r}^{3}$ for $\left|\Psi_{0}\right|^{2}$.
}
\end{minipage}
\end{figure}

\begin{figure}
\centering
\begin{minipage}[b]{1\linewidth}
\includegraphics[width=1\textwidth]{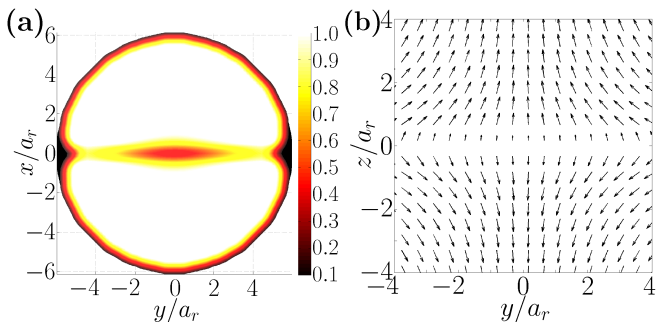}
 \caption{\label{fig:spinkuvat}
(a) Spin density for the strong-field seeking ground-state monopole in the $z=0$ plane. Spin density is depleted 
along the two vortices, which lie on the $y$-axis. The figure corresponds to spin densities 
from $0.1\times{10^{-4}N/a_{r}^{3}}$ to $0.92\times{10^{-4}N/a_{r}^{3}}$.
(b) Spin of the ground-state monopole. The arrows represent the projection  of the spin
to a plane corresponding to $x=0$. The $S_{x}$ component is zero in this plane.
}
\end{minipage}
\end{figure}

In the WFSS, the Dirac strings lie on the $z$-axis because
the magnetic trap is strongest along this direction. For the SFSS, the confinement
in the $z$-direction is weaker than in the $x$- and $y$-directions 
due to the symmetric optical trap, and 
the two strings lie in the $xy$-plane. For the SFSS, we also consider an asymmetric
optical trap of the form
$V_{\textrm{opt}}\left(\boldsymbol{r}\right)=m\left(\omega_{x}^{2}x^{2}+\omega_{y}^{2}y^{2}
+\omega_{z}^{2}z^{2}\right)/2$. In general, the Dirac strings associated with the monopole 
defect lie along the direction that minimizes the length of the vortex lines. Particle densities 
for the SFSS and WFSS are shown in Fig.~\ref{fig:dens}. For the WFSS, the vortices 
corresponding to the Dirac strings are singular and the 
particle density vanishes at the vortex core.
In the case of the SFSS, particle density is only partially depleted along the two
vortices, implying that they are polar-core vortices~\cite{Isoshima:2001}. The particle densities for
each spin state in the WFSS and SFSS are shown in Fig.~\ref{fig:spinor}.
In the SFSS, the two vortices are manifested as depletion in the spin density, see 
Fig.~\ref{fig:spinkuvat}(a).

\begin{figure}
\begin{minipage}[b]{1\linewidth}
\includegraphics[width=1\textwidth]{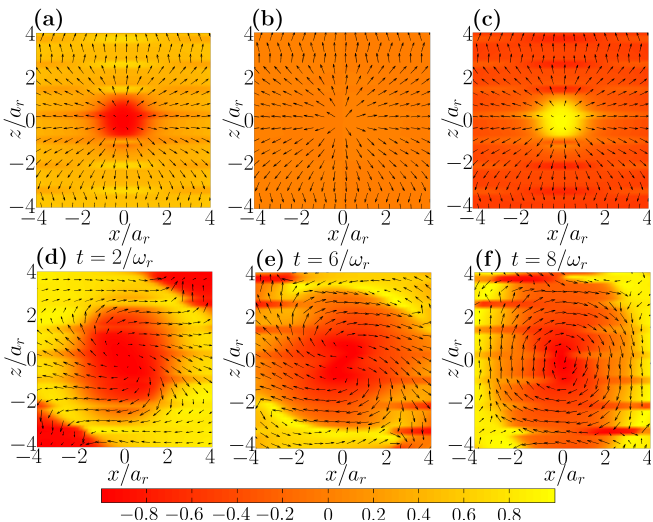}
\caption{\label{fig:vorticity}(a)--(c) Vorticity of the ground-state monopole at different locations.
(d)--(f) Temporal evolution of vorticity at fixed location. In panels (a)--(c), arrows represent a
projection of the unit vorticity $\hat{\Omega}_{s}=\boldsymbol{\Omega}_{s}/\left|\boldsymbol{\Omega}_{s}\right|$ to
$xz$-plane for (a) $y=0.8\times{a_{r}}$, (b) $y=0\times{a_{r}}$ and (c) $y=-0.8\times{a_{r}}$.
In panels (d)--(f) $y=-0.7\times{a_{r}}$ at time instants (d) $t=2/\omega_{r}$, (e) $t=6/\omega_{r}$ 
and (f) $t=8/\omega_{r}$. The $y$ component is presented with the colormap.}
\end{minipage}
\end{figure}

Vorticities and spin textures corresponding to SFSS and WFSS were
found to be qualitatively similar, and we only present further results for the SFSS.
The condensate spin 
is shown in Fig.~\ref{fig:spinkuvat}(b) and one observes that the
spin indeed tends to align with the local magnetic field. From Figs.~\ref{fig:vorticity}(a)--(c) 
we observe that vorticity 
takes the hedgehog form of Fig.~\ref{fig:textures}(b) near the monopole defect.

\begin{figure}[!h]
\begin{minipage}[b]{1\linewidth}
\includegraphics[width=0.9\textwidth]{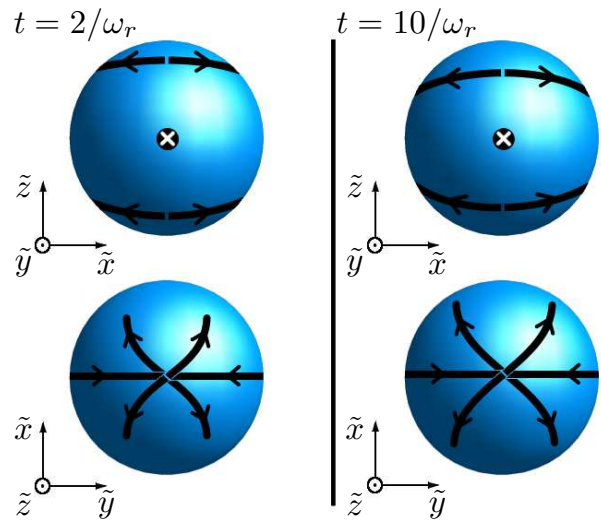}
\caption{\label{fig:schematic}
{Schematic representation of the temporal evolution of the monopole defect.} The spheres represent the atomic 
cloud and the black lines correspond to vortices. The arrows illustrate the direction of vorticity associated 
with the single-quantum vortices. The condensate is depicted at two instants of time $t=2/\omega_{r}$ and
$t=10/\omega_{r}$. In the top panels the condensate is viewed along the $y$-axis and in the bottom panels,
along the $z$-axis.
}
\end{minipage}
\end{figure}

Next we study the dynamics of the monopole after the external  magnetic fields are
turned off. We take the initial state to be the strong-field seeking ground state with vortices on
the $y$-axis. The fields are turned off instantaneously and the state is evolved up to 
$10/\omega_{r}$ with a timestep $10^{-4}/\omega_{r}$. During the time evolution, 
the Dirac strings are observed to expand,
see Figs.~\ref{fig:vorticity}(d)--(f).
Furthermore,
two antimonopoles emerge from the boundary of the condensate and move 
towards the center of the trap.
The emergent antimonopoles are characterized by Dirac strings that carry vorticity outwards 
from the monopole. Similarly to the ground-state monopole, the Dirac strings are coreless vortices.
We note that the Dirac strings associated with antimonopoles guarantee that the total vorticity
remains zero after the antimonopoles emerge. A schematic representation of the temporal evolution
is shown in Fig.~\ref{fig:schematic} and high resolution
figures of the simulation are available in the supporting online material~\cite{EPAPS}.
At the end of the simulation, the vorticity
of the monopole defect has spread out and is no longer strictly of the hedgehog form.

In conclusion, we have shown that a monopole defect may exist as the ground state of a dilute ferromagnetic spin-1
BEC in the presence of an experimentally feasible magnetic field configuration. Vorticity of this defect is analogous
to the magnetic field of a magnetic monopole. The ground-state monopole is associated with two Dirac strings that extend 
symmetrically outwards from the monopole defect.
From the relation between the geometric charge of the monopole defect and the total vorticity 
in the system, it follows that each of the two coreless vortices associated with the SFSS 
monopole defect carry a single quantum of angular momentum. This is remarkable, since in 
general the angular momentum associated with coreless vortices is not quantized. 
Thus, the monopole defect is responsible for the quantization of coreless vortices similarly to 
the Dirac magnetic monopole which imposes the quantization of electric charge.  
If monopoles are to appear spontaneously in BECs, one would expect for energetic reasons that they are  
associated with two singly quantized vortices instead of a single two-quantum vortex. 
Hence such monopoles are of great interest. In the experiments, the ground-state monopole 
can be created during the cooling process in situ, without the need to adjust the 
external magnetic fields in time. These monopoles are expected to be extremely robust and long lived.

\begin{acknowledgments}
We acknowledge the Emil Aaltonen Foundation, the Academy of Finland, Finnish Graduate School 
in Computational Sciences, The Research Foundation of Helsinki University of Technology, 
and Harvard--MIT CUA for financial support and the Center for Scientific Computing, 
Finland for computing resources.
\end{acknowledgments}
\bibliography{monopole}
\bibliographystyle{apsrev4-1}

\end{document}